\documentclass[12pt]{article}
\usepackage{latexsym}
\usepackage{graphicx}
\usepackage{epsfig}
\newcommand \be {\begin{equation}}
\newcommand \bea {\begin{eqnarray}}
\newcommand \ee {\end{equation}}
\newcommand \eea {\end{eqnarray}}

\newcommand{\bit}{\begin{itemize}}
\newcommand{\eit}{\end{itemize}}

\newcommand{\eps}{\epsilon}

 \renewcommand{\theequation}{\thesection.\arabic{equation}}
 \def\appendixa{
 \vskip 1cm
 \noindent
 {\Large \bf Appendix A: Construction of the General Relativistic}
 \vskip 0.5cm
 \noindent {\Large \bf Kink}
 \vskip 1cm
 \par
 \setcounter{equation}{0}
 \def\theequation{A.\arabic{equation}}
 }
 \def\appendixb{
 \vskip 1cm
 \noindent
 {\Large \bf Appendix B: Flat vs General Relativistic Kinks}
 \vskip 1cm
 \par
 \setcounter{equation}{0}
 \def\theequation{B.\arabic{equation}}
 }
\begin{document}
\topskip 2cm

\begin{titlepage}


\begin{center}
{\Large\bf Dimensionally Reduced Gravitational }\\

\vspace{0.5cm}

{\Large \bf Chern-Simons Term and its Kink}

\vspace{2.5cm}

{\large G. Guralnik$^{1, *}$, A. Iorio$^{1,2}$, R. Jackiw$^{1}$, S.-Y. Pi$^{3}$} \\
\vspace{.5cm}
{\sl $^{1}$ Center for Theoretical Physics, Massachusetts Institute of Technology}\\
{\sl Cambridge MA 02139 USA} \\

{\sl $^{2}$ Istituto Nazionale di Fisica Nucleare, Rome, Italy} \\

{\sl $^{3}$ Physics Department, Boston University, Boston MA 02215 USA}\\

\vspace{2.5cm}

\begin{abstract}
\noindent When the gravitational Chern-Simons term is reduced from
3 to 2 dimensions, the lower dimensional theory supports a
symmetry breaking solution and an associated kink. Kinks in
general relativity bear a close relation to flat space kinks,
governed by identical potentials.
\end{abstract}
\end{center}

\vspace{1.5cm}

\noindent MIT-CTP-3373, BUHEP 03-11, BROWNHET-1357

\noindent hep-th/0305117

\vfill

\noindent ${}^*$ On leave of absence from Physics Department,
Brown University, Providence RI 02912, USA.

\end{titlepage}

\newpage

\section{Introduction}

Reducing the Hilbert-Einstein gravity action from 3- to 2-dimensional
space-time produces an interesting 2-dimensional
(lineal) gravity theory \cite{jt}, which has engendered much
further research \cite{kummer}. But in 3 dimensions there exists
another invariant action: the gravitational Chern-Simons term
\cite{djt}, whose reduction to 2 dimensions is the subject of this
paper. The present work is a continuation of earlier
investigations, wherein Lie-group-based Chern-Simons terms are
reduced in various ways \cite{jp}.

In Section 2, we recall the construction of the gravitational
Chern-Simons term. In Section 3, a Kaluza-Klein {\it Ansatz} is
posited for the 3-metric tensor, and the 3-dimensional
Chern-Simons action is reduced to 2 dimensions. There it retains
some topological features and is reminiscent of an
axion/Chern-Pontryagin structure. Some properties of the emergent
lineal gravity theory are explored in Section 4. In particular,
attention is called to a symmetry breaking solution, which gives
rise to anti-de Sitter space. Also, there is a kink solution that
deforms the anti-de Sitter space as it interpolates between the
symmetry breaking configurations: asymptotically reproducing the
anti-de Sitter space, but possessing positive curvature in finite
regions. Concluding remarks comprise Section 5. Two Appendices are
devoted to detailed calculations. In Appendix A we give the
derivation of our general relativistic kink. In Appendix B a
relation is demonstrated between flat space and general
relativistic kinks, governed by the same potential.

\section{Gravitational Chern-Simons Term}

The Chern-Simons term, based on a Lie algebra, with matrix-valued
gauge connections $A_\mu$, takes the form
\begin{equation}\label{csa}
{\rm CS} (A) = \frac{1}{4 \pi^2} \int d^3 x \eps^{\mu \nu \lambda}
{\rm tr} \left( \frac{1}{2} A_\mu \partial_\nu A_\lambda +
\frac{1}{3} A_\mu A_\nu A_\lambda \right) \;.
\end{equation}
Its variation with respect to $A_\mu$ leads to the dual ${}^*
F^\mu$ of the curvature $F_{\mu \nu}$
\begin{eqnarray}
  F_{\mu \nu} &=& \partial_\mu A_\nu - \partial_\nu A_\mu + [A_\mu, A_\nu] \;,\\
 {}^* F^\mu &=& \frac{1}{2} \eps^{\mu \nu \lambda} F_{\nu \lambda} \;,\label{dualf}
\end{eqnarray}
\begin{equation}\label{deltacsa}
\delta {\rm CS} (A) = \frac{1}{4 \pi^2} \int d^3 x \; {\rm tr} \left(
 {}^* F^\mu \delta A_\mu \right) \;.
\end{equation}
While (\ref{csa}) is invariant against infinitesimal gauge
transformations, finite gauge transformations
\begin{equation}\label{atoaprime}
  A_\mu \to U^{-1} A_\mu U + U^{-1} \partial_\mu U \;,
\end{equation}
produce the response
\begin{equation}\label{csatocsaprime}
{\rm CS} (A) \to {\rm CS} (A) - \frac{1}{24 \pi^2} \int d^3 x
\eps^{\mu \nu \lambda} {\rm tr} \left( U^{-1} \partial_\mu U
U^{-1} \partial_\nu U U^{-1} \partial_\lambda U \right) \;,
\end{equation}
apart from a surface integral, which we drop. The last term in
(\ref{csatocsaprime}) is the ``winding number" $W(U)$ of the gauge
transformation function $U$. Even though the variation of $W(U)$
is a surface term
\begin{equation}\label{deltaw}
  \delta W(U) = - \frac{1}{8 \pi^2} \int d^3 x
\eps^{\mu \nu \lambda} \partial_\mu {\rm tr} \left( U^{-1}
\partial_\nu U U^{-1} \partial_\lambda U U^{-1} \delta U
\right) \;,
\end{equation}
$W(U)$ can be non vanishing for homotopically non trivial $U$.
This may require that the strength of the Chern-Simons term be
quantized \cite{djt}.

Analogous properties are enjoyed by the gravitational Chern-Simons
term. This is made explicit when we use an adapted notation,
defined as follows.

Consider the Christoffel connection, constructed from the metric
tensor\footnote{In our notation, capital letters denote quantities
in the larger space, here 3-dimensional, while lower case is used
for corresponding quantities in the reduced space, here
2-dimensional. Also, letters from the middle of the Greek alphabet
($\lambda, \mu, \nu, ...$) refer to space-time components in the
larger space, while beginning letters ($\alpha, \beta, \gamma,
...$) denote components in the reduced, 2-dimensional space.
Finally, tangent space (flat space) components, to be introduced
below, are described by Latin letters, capitalized in 3-space,
lower-case in 2-space.} $G_{\mu \nu}$
\begin{equation}\label{Chris}
  \Gamma^\lambda_{\mu \nu} = \frac{1}{2} G^{\lambda \sigma} \left(
  \partial_\mu G_{\sigma \nu} + \partial_\nu G_{\sigma \mu} -
  \partial_\sigma G_{\mu \nu} \right) \;.
\end{equation}
The coordinate transformation law for $\Gamma^\lambda_{\mu \nu}$
\begin{equation}\label{ChristoChrisprime}
  \Gamma^\lambda_{\mu \nu} (x) \to \tilde{\Gamma}^\lambda_{\mu \nu}
  (\tilde{x}) = \frac{\partial \tilde{x}^\lambda}{\partial x^\rho}
  \frac{\partial {x}^\sigma}{\partial \tilde{x}^\mu}
  \frac{\partial {x}^\tau}{\partial \tilde{x}^\nu}
  \Gamma^\rho_{\sigma \tau} (x) +
  \frac{\partial \tilde{x}^\lambda}{\partial x^\rho}
  \frac{\partial^2 {x}^\rho}{\partial \tilde{x}^\mu \partial
  \tilde{x}^\nu}\;,
\end{equation}
may be succinctly presented in the following manner. Consider
$\Gamma^\lambda_{\mu \nu}$ to be a matrix valued connection with
space-time index $\mu$, and matrix indices ($\lambda, \nu$)
\begin{equation}\label{ChirAmu}
  \Gamma^\lambda_{\mu \nu} = \left( A_\mu \right)^\lambda_{\;\;\nu} \;.
\end{equation}
Then (\ref{ChristoChrisprime}) is recognized as a coordinate
transformation on the index $\mu$, and a gauge transformation
(\ref{atoaprime}) with gauge function
\begin{equation}\label{ugrav}
  \left( U \right)^\tau_{\;\; \nu} = \frac{\partial {x}^\tau}{\partial
  \tilde{x}^\nu} \;, \quad
  \left( U^{-1} \right)^\lambda_{\;\; \rho} = \frac{\partial \tilde{x}^\lambda}{\partial
  {x}^\rho} \;,
\end{equation}
\begin{equation}\label{atouau}
\left( A_\mu (x) \right)^\lambda_{\;\; \nu}  \to  \left(
\tilde{A}_\mu (\tilde{x}) \right)^\lambda_{\;\; \nu}
  = \frac{\partial {x}^\sigma}{\partial
  \tilde{x}^\mu} \left( U ^{-1} A_\sigma (x) U + U^{-1} \frac{\partial }{\partial
  {x}^\sigma} U \right)^\lambda_{\;\;\; \nu}  \;.
\end{equation}
Moreover, the formula for the Riemann curvature tensor
\begin{equation}\label{riemann}
  R^\rho_{\; \sigma \mu \nu} = \frac{\partial \Gamma^\rho_{\mu \sigma}}{\partial
  {x}^\nu} - \frac{\partial \Gamma^\rho_{\nu \sigma}}{\partial
  {x}^\mu} + \Gamma^\rho_{\nu \tau} \Gamma^\tau_{\mu \sigma} -
  \Gamma^\rho_{\mu \tau} \Gamma^\tau_{\nu \sigma} \;,
\end{equation}
coincides with the gauge theoretic curvature of $A_\mu$
\begin{equation}\label{r=f}
R^\rho_{\; \sigma \mu \nu} = \left( F_{\nu \mu} \right)^\rho_{\;\;
\sigma} = \left( \partial_\nu A_\mu -
\partial_\mu A_\nu + [A_\nu, A_\mu] \right)^\rho_{\;\; \sigma} \;.
\end{equation}

The {\it Vielbein}/spin connection formulation of gravity is also
usefully presented with the above variables. Denoting the {\it
Vielbein} by $E^A_{\;\; \mu}$ and its inverse by\footnote{Here
$\eta = {\rm diag} (+1,-1, ...)$ is the tangent space (flat space)
metric tensor.} $E^\mu_{\;\; A} = G^{\mu \nu} \eta_{A B} E^B_{\;\;
\nu}$, we have the defining property
\begin{equation}\label{vielbein}
G_{\mu \nu} = \eta_{A B} E^A_{\;\; \mu} E^B_{\;\; \nu} \;.
\end{equation}
The metricity condition defines the spin connection $\left(
\Omega_\mu \right)^A_{\;\; B}$
\begin{equation}\label{derviel}
  D_\mu E^A_{\;\; \nu} = \partial_\mu E^A_{\;\; \nu} - \Gamma^\lambda_{\mu \nu}
  E^A_{\;\; \lambda} = - \left( \Omega_\mu \right)^A_{\;\; B} E^B_{\;\; \nu} \;.
\end{equation}
But with (\ref{ChirAmu}) this may be rewritten as
\begin{equation}\label{AmuOmega}
  \left( A_\mu \right)^\lambda_{\;\; \nu} = E_{\;\; A}^\lambda \left( \Omega_\mu \right)^A_{\;\; B}
  E^B_{\;\; \nu} + E_{\;\; A}^\lambda \partial_\mu E^A_{\;\; \nu} \;,
\end{equation}
showing that the Christoffel connection $A_\mu$ [matrix indices
$(\lambda , \nu)$] is the gauge transform of the spin connection
$\Omega_\mu$ [matrix indices $(A,B)$] with the gauge function
$U^B_{\;\; \nu} = E^B_{\;\; \nu}$. It then follows immediately
from (\ref{r=f}) that the Riemann curvature tensor constructed
from the spin connection
\begin{equation}\label{r=fomega}
  R^A_{\;\; B \mu \nu} = \left( \partial_\nu \Omega_\mu - \partial_\mu
\Omega_\nu + [\Omega_\nu, \Omega_\mu] \right)^A_{\;\; B} \;,
\end{equation}
is related to (\ref{r=f}) by
\begin{equation}
R^\rho_{\; \sigma \mu \nu} = E^\rho_{\;\; A} R^A_{\;\; B \mu \nu}
E^B_{\;\; \sigma} \;.
\end{equation}

This notation provides a gauge theory/gravity theory dictionary,
with which we can translate the gauge theoretic Chern-Simons term
(\ref{csa}) into the gravitational one in terms of the Christoffel
connection. Using (\ref{ChirAmu}) in (\ref{csa}) leads to
\begin{equation}\label{cschris}
{\rm CS} (\Gamma) = \frac{1}{4 \pi^2} \int d^3 x \eps^{\mu \nu
\lambda} \left( \frac{1}{2} \Gamma^\rho_{\mu \sigma} \partial_\nu
\Gamma^\sigma_{\lambda \rho} + \frac{1}{3} \Gamma^\rho_{\mu
\sigma} \Gamma^\sigma_{\nu \tau} \Gamma^\tau_{\lambda \rho}
\right) \;.
\end{equation}
Moreover, (\ref{atoaprime}), (\ref{csatocsaprime}),
(\ref{ChirAmu}), and (\ref{AmuOmega}) provide an alternative
formula for CS($\Gamma$) in terms of the spin connection
\begin{equation}\label{csOmega}
{\rm CS} (\Gamma) = \frac{1}{4 \pi^2} \int d^3 x \eps^{\mu \nu
\lambda} {\rm tr} \left( \frac{1}{2} \Omega_{\mu}
\partial_\nu \Omega_{\lambda} + \frac{1}{3}
\Omega_{\mu} \Omega_{\nu} \Omega_{\lambda} \right) - \frac{1}{24
\pi^2} \int d^3 x \eps^{\mu \nu \lambda} {\rm tr} \left(V_{\mu}
V_{\nu} V_{\lambda} \right) \;.
\end{equation}
The matrix $(V_\mu)^{\sigma}_{\;\; \rho}$ is seen from
(\ref{AmuOmega}) to be $E^\sigma_{\;\; A} \partial_\mu E^A_{\;\;
\rho}$; thus the last term is the winding number $W(E)$ of the
{\it Dreibein} $E^A_{\;\; \mu}$.

According to (\ref{deltaw}), the variation of $W(E)$ is a surface
term, therefore variations of CS($\Gamma$) coincide with those of
CS($\Omega$) -- the next-to-last contribution to (\ref{csOmega})
-- and $W(E)$ does not contribute to the equations of motion.
However, invariance against local rotations -- the gauge group of
the spin connection -- is assured only when $W(E)$ is kept: the
non-invariance of CS($\Omega$) [c.f. (\ref{csatocsaprime})] is
compensated by the non-invariance of $W(E)$. This is as it must
be, because the two together reproduce CS($\Gamma$), which does
not respond to local rotations. Thus, regardless of the homotopy
properties of local rotations, there are no quantization
requirements on the gravitational Chern-Simons coefficient.

Correspondingly, diffeomorphisms preserve the world scalar
CS($\Omega$). One can check that the non invariance of
CS($\Gamma$) in (\ref{csOmega}) is carried by the last term $W(E)$
which changes as in (\ref{csatocsaprime}), with $U$ given by
(\ref{ugrav})\footnote{This analysis of the invariance properties
of the gravitational Chern-Simons term was first performed by
R.Percacci \cite{percacci}, who also argued that diffeomorphisms
(orientation preserving, and tending to the identity at infinity)
do not possess a non-vanishing winding number. Therefore no
quantization of the Chern-Simons coefficient is needed in the
metric formulation, consistent with the conclusion found in the
spin connection formulation.}.

We note that in 3 dimensions the antisymmetric quantity
$\Omega_{\mu \;, A B} = \eta_{A C} (\Omega_\mu)^C_{\;\;\ B}$ may
be written as $\eps_{A B C} \Omega^C_{\;\; \mu}$. Consequently a
simpler expression for CS($\Omega$) is available
\begin{eqnarray}\label{csOmegavera}
{\rm CS} (\Omega) &=& - \frac{1}{4 \pi^2} \int d^3 x \eps^{\mu \nu
\lambda} \left( \eta_{A B}\;  \Omega^A_{\; \; \mu} \partial_\nu
\Omega^B_{\; \; \lambda} - \frac{1}{3} \eps_{A B C} \Omega^A_{\;
\;  \mu} \Omega^B_{\; \; \nu} \Omega^C_{\; \; \lambda}
\right) \nonumber \\
&=& - \frac{1}{4 \pi^2} \int d^3 x \eps^{\mu \nu \lambda} \eta_{A
B} \Omega^A_{\; \; \mu} \partial_\nu \Omega^B_{\; \; \lambda} +
\frac{1}{2 \pi^2} \int d^3 x \; {\rm det} \; \Omega^A_{\; \; \mu}
\;.
\end{eqnarray}

The variation of the gravitational Chern-Simons term determines
the Cotton tensor. The derivation is facilitated by our gauge
theoretical dictionary. From (\ref{deltacsa}), (\ref{ChirAmu}),
and (\ref{r=f}) we have
\begin{equation}\label{deltacsgamma}
\delta {\rm CS} (\Gamma) = \frac{1}{8 \pi^2} \int d^3 x \eps^{\mu
\nu \lambda } R^\rho_{\; \sigma \lambda \nu} \delta
\Gamma^\sigma_{\mu \rho} \;.
\end{equation}
Next we use the 3-dimensional formula for the Riemann tensor in
terms of the Ricci tensors and scalar curvature
\begin{equation}\label{riemann2+1}
  R^\rho_{\; \sigma \lambda \nu} = \delta_\lambda^\rho (R_{\sigma \nu} -
  \frac{1}{2} G_{\sigma \nu} R) - \delta_\nu^\rho (R_{\sigma \lambda} -
  \frac{1}{2} G_{\sigma \lambda} R) + G_{\sigma \nu}
  R^\rho_\lambda - G_{\sigma \lambda} R^\rho_\nu \;,
\end{equation}
and the expression for the variation of the Christoffel connection
\begin{equation}\label{deltaGamma}
  \delta \Gamma^\sigma_{\mu \rho} = \frac{1}{2} G^{\sigma \tau}
  (D_\mu \delta G_{\tau \rho} + D_\rho \delta G_{\tau \mu} -
  D_\tau \delta G_{\mu \rho}) \;,
\end{equation}
to write (\ref{deltacsgamma}), after a partial integration, as
\begin{eqnarray}
\delta {\rm CS} (\Gamma) &=& \frac{1}{4 \pi^2} \int d^3 x
\eps^{\mu \nu \sigma } ( D_\mu R^\tau_{\nu}) \delta G_{\sigma
\tau} \nonumber \\
&\equiv& - \frac{1}{8 \pi^2}  \int d^3 x \; \sqrt{G} C^{\sigma
\tau} \delta G_{\sigma \tau} \label{deltacscotton} \;,
\end{eqnarray}
where $G = {\rm det} \; G_{\mu \nu}$, and the Cotton tensor
$C^{\sigma \tau}$ is
\begin{equation}\label{cotton}
C^{\sigma \tau} = - \frac{1}{2} \frac{1}{\sqrt{G}} (\eps^{\mu \nu
\sigma } D_\mu R^\tau_{\nu} + \eps^{\mu \nu \tau } D_\mu
R^\sigma_{\nu}) \;.
\end{equation}
This is traceless, indicating that CS($\Gamma$) is conformally
invariant.

\section{Dimensional Reduction}

To effect the dimensional reduction from 3 ($t, x, y$) to 2 ($t,
x$) dimensions, we choose the Lorentz signature and set a
Kaluza-Klein {\it Ansatz} for the metric tensor $G_{\mu \nu}$
\cite{bergmann}
\begin{equation}\label{Gred}
  G_{\mu \nu} = \phi \left( \begin{array}{ccc}
    g_{\alpha \beta} - a_\alpha a_\beta &  & - a_\alpha \\
    - a_\beta &  & - 1 \
  \end{array}\right) \;.
\end{equation}
Here ($\mu,\nu = 0,1,2$), $g_{\alpha \beta}$ is the metric tensor
for our 2-space, $a_\alpha$ is a 2-vector, ($\alpha, \beta =
0,1$), and $\phi$ is a scalar. All quantities depend on the first
two coordinates ($t, x$),  and are independent of the third
coordinate $y$. When a general $y$-independent coordinate
transformation, $\delta x^\mu = - \xi^\mu (t,x) $, is performed on
$G_{\mu \nu}$
\begin{equation}\label{deltaG}
  \delta G_{\mu \nu} = \xi^\rho \partial_\rho G_{\mu \nu} +
  \partial_\mu \xi^\rho G_{\rho \nu} + \partial_\nu \xi^\rho G_{\mu
  \rho} \;,
\end{equation}
it follows that
\begin{eqnarray}
- \delta G_{2 2} & = & \delta \phi = \xi^\alpha
\partial_\alpha \phi \;, \nonumber \\
- \delta ( G_{\alpha 2} / \phi ) & = & \delta a_\alpha = \xi^\beta
\partial_\beta a_\alpha + \partial_\alpha \xi^\beta
a_\beta + \partial_\alpha \xi^2 \;, \label{deltaGcomp}\\
\delta ( [ G_{\alpha \beta} + a_\alpha a_\beta ] / \phi ) & = &
\delta g_{\alpha \beta} = \xi^\gamma \partial_\gamma g_{\alpha
\beta} + \partial_\alpha \xi^\gamma g_{\gamma \beta} +
\partial_\beta  \xi^\gamma g_{\alpha \gamma} \nonumber
\;.
\end{eqnarray}
This shows that the reduced quantities transform properly under
2-dimensional diffeomorphisms, and also that $a_\alpha$ undergoes
an Abelian gauge transformation with gauge function $\xi^2$.
Consequently the reduced action will be a gauge and coordinate
invariant. Moreover, because the Chern-Simons term is conformally
invariant, it will not depend on $\phi$, which we henceforth set
to 1. With (\ref{Gred}) and $\phi=1$, the components of the
Christoffel connection are
\begin{eqnarray}
\Gamma^\alpha_{\beta \gamma} & = & \gamma^\alpha_{\beta \gamma} -
g^{\alpha \delta} (a_\beta f_{\gamma \delta} + a_\gamma f_{\beta
\delta }) \;, \nonumber \\
\Gamma^2_{\beta \gamma} & = & \frac{1}{2} (D_\beta a_\gamma +
D_\gamma a_\beta) + \frac{1}{2} a^\delta (a_\beta f_{\gamma
\delta} + a_\gamma f_{\beta \delta}) \;, \label{Chrisred} \\
\Gamma^\alpha_{2 \beta} & = & \frac{1}{2} g^{\alpha \gamma}
f_{\gamma \beta} \;, \quad \Gamma^2_{2 \gamma} = \frac{1}{2}
f_{\gamma \delta} a^{\delta} \;, \quad \Gamma^\alpha_{2 2} =
\Gamma^2_{2 2} = 0 \nonumber \;.
\end{eqnarray}
The inverse of $g_{\alpha \beta}$ is $g^{\alpha \beta}$, and all
indices are moved with this 2-dimensional metric tensor. The
Abelian field strength $f_{\alpha \beta}$ is constructed from
$a_\alpha$: $f_{\alpha \beta} = \partial_{\alpha} a_\beta -
\partial_{\beta} a_\alpha$.  $D$ is the covariant derivative based
on the connection $\gamma^\alpha_{\; \beta \gamma}$.

For the {\it Dreibein} $E^A_{\;\;\ \mu}$ and spin connection
$\Omega^A_{\;\;\ \mu}$ we have
\begin{equation}\label{2bein}
  E^a_{\;\; \alpha} = e^a_{\;\; \alpha} \;, \; E^2_{\;\; \alpha} = a_\alpha \;, \;
  E^a_{\;\; 2} = 0 \;, \; E^2_{\;\; 2} = 1 \;,
\end{equation}
and
\begin{equation}\label{2spinconn}
   \Omega^a_{\;\; \alpha} =  \frac{1}{2} e^a_{\;\; \alpha} f
 \;, \; \Omega^2_{\;\; \alpha} = - \omega_\alpha - \frac{1}{2} f a_\alpha
\;, \; \Omega^a_{\;\; 2} = 0 \;, \; \Omega^2_{\;\; 2} = -
\frac{1}{2} f \;.
\end{equation}
Here $e^a_{\; \alpha}$ is the {\it Zweibein}, the 2-dimensional
spin connection $(\omega_ \alpha)^a_{\; b}$ is written in terms of
2-dimensional antisymmetric symbol as $\omega_{\alpha , \; a b} =
\eps_{a b} \; \omega_\alpha$, while $f$ is the invariant field
strength
\begin{equation}\label{f}
  f_{\alpha \beta} =  \sqrt{- g} \; \eps_{\alpha \beta} f
  \;,
\end{equation}
$g = {\rm det} \; g_{\alpha \beta}$, $\eps^{0 1} = 1$. With these,
we find for the gravitational Chern-Simons term (apart from
surface contributions)
\begin{equation}\label{cspiccola}
  {\rm CS}  = - \frac{1}{8 \pi^2} \int d^2 x \sqrt{-g} (f r + f^3)
  \;,
\end{equation}
where $r$ is the 2-dimensional scalar curvature. Note that the
topological origin of this expression is recognized when we
express $\sqrt{-g} f d^2 x$ as the 2-form $d a$, with $a =
a_\alpha d x^\alpha$. Thus we also have
\begin{equation}\label{cspiccolaform}
  {\rm CS}  = \frac{1}{8 \pi^2} \int da (r + f^2)
  \;.
\end{equation}
Another intriguing rewriting presents (\ref{cspiccolaform}) as
\begin{equation}\label{cspiccolaxion}
  {\rm CS}  = \frac{1}{4 \pi} \int d^2 x \; \Theta \; \eps^{\alpha \beta} f_{\alpha
  \beta}\;,
\end{equation}
where $\Theta = \frac{1}{4 \pi} (r + f^2)$. This puts into
evidence an axion ($\Theta$) - Chern-Pontryagin ($\eps^{\alpha
\beta} f_{\alpha \beta}$ in 2 dimensions) structure. Finally, we
remark that the 3-dimensional scalar curvature $R$ with {\it
Ansatz} (\ref{Gred}) (and $\phi = 1$) reduces to \cite{bergmann}
\begin{equation}\label{Rred}
  R = r + \frac{1}{2} f^2 \;.
\end{equation}

\section{Variations and Equations of Motion}

Variation of (\ref{cspiccolaform}) produces
\begin{equation}\label{deltacsred}
  \delta {\rm CS} = \frac{1}{4 \pi^2} \int d^2 x \sqrt{- g}
  (- j^\alpha \delta a_\alpha + \frac{1}{2} T_{\alpha \beta} \delta g^{\alpha
  \beta})\;,
\end{equation}
where
\begin{eqnarray}
j^\alpha & = & - \frac{1}{2 \sqrt{-g}} \eps^{\alpha \beta}
\partial_\beta (r + 3 f^2) \;, \label{j} \\
T_{\alpha \beta} & = & g_{\alpha \beta} (D^2 f - f^3 - \frac{1}{2}
r f) - D_\alpha D_\beta f \label{T} \;.
\end{eqnarray}
Note that identically $D_\alpha j^\alpha = 0$, as consequence of
gauge invariance. For the second rank tensor we have
\begin{equation}\label{DT1}
  D^\beta T_{\alpha \beta} = - \partial_{\alpha} (f^3 + \frac{1}{2} r f)
  + [D_\alpha , D_\beta] D^\beta f \;.
\end{equation}
The last term is $\frac{1}{2} r \partial_\alpha f$, so (\ref{DT1})
becomes
\begin{equation}\label{DT2}
  D^\beta T_{\alpha \beta} = - \frac{1}{2} f \partial_{\alpha} (r + 3 f^2)
  = j^\beta f_{\beta \alpha} \;,
\end{equation}
which vanishes with $j^\beta$, as it should as consequence of
2-dimensional diffeomorphism invariance.

The components of the dimensionally reduced Cotton tensor are
\begin{equation}\label{cottonred}
C^{\alpha \beta} = T^{\alpha \beta} \;,\; C^{\alpha 2} = -
j^\alpha - T^{\alpha \beta} a_\beta  \;,\; C^{22} = g_{\alpha
\beta } T^{\alpha \beta} + a_\alpha T^{\alpha \beta} a_\beta + 2
j^\alpha a_\alpha \;.
\end{equation}

If the entire dynamics is governed by the Chern-Simons term, the
equations of motion become
\begin{eqnarray}
0 & = & \eps^{\alpha \beta}
\partial_\beta (r + 3 f^2) \;, \label{elj} \\
0 & = & g_{\alpha \beta} (D^2 f - f^3 - \frac{1}{2} r f) -
D_\alpha D_\beta f \label{elT} \;.
\end{eqnarray}
The first is solved by
\begin{equation}\label{eljsol}
  r + 3 f^2 = {\rm constant} = c \;.
\end{equation}
Eliminating $r$ in (\ref{elT}), and decomposing into the trace and
trace-free parts lead to
\begin{eqnarray}
0 & = & D^2 f - c f + f^3 \;, \label{elf1} \\
0 & = & D_\alpha D_\beta f - \frac{1}{2} g_{\alpha \beta} D^2 f
\label{elf2} \;.
\end{eqnarray}
Note that the equations are invariant against changing the sign of
$f$ (the action changes sign). A solution that respects this
``symmetry'' is
\begin{equation}\label{symsol}
  f = 0 \;,\; r = c \;.
\end{equation}
However, there is also a "symmetry breaking" solution
\begin{equation}\label{symbreksol}
  f = \pm \sqrt{c} \;,\; r = - 2 c \;, \quad \quad c > 0\;.
\end{equation}

The 3-dimensional curvature associated with the above two
solutions is, according to (\ref{Rred}), $R=c$ for the symmetry
preserving case (\ref{symsol}), and $R= - \frac{3}{2} c$ ($c > 0$)
for the symmetry breaking one. Although both show constant
curvature, the latter is maximally symmetric in the 3-dimensional
geometrical sense, i.e. one verifies that the 3-dimensional Ricci
tensor $R_{\mu \nu}$ is given by $\frac{1}{3} g_{\mu \nu} R$, with
six Killing vectors that span $SO(2,1) \times SO(2,1) = SO(2,2)$,
the isometry of 3-dimensional anti-de Sitter space. The former is
not maximally symmetric, and it admits only four Killing vectors
that span $SO(2,1) \times SO(2)$.

When $c > 0$, so that the symmetry breaking solution
(\ref{symbreksol}) is present, there also is a kink solution. The
profile for $f$
\begin{equation}\label{fkink}
  f = \sqrt{c} \tanh \frac{\sqrt c}{2} x \;,
\end{equation}
interpolates between the two homogeneous solutions in
(\ref{symbreksol}), and with (\ref{eljsol}) gives rise to the
2-dimensional curvature
\begin{equation}\label{rkink}
  r = - 2 c + 3 c \frac{1}{\cosh^2 \frac{\sqrt c}{2} x} \;.
\end{equation}
The relevant line element is
\begin{equation}\label{dskink}
  \left( d s \right)^2 = \frac{1}{\cosh^4 \frac{\sqrt c}{2} x}
  \left( d t \right)^2 - \left( d x \right)^2 \;.
\end{equation}
One may directly verify that (\ref{eljsol})--(\ref{elf2}) are
satisfied by (\ref{fkink})--(\ref{dskink}). In Appendix A we give
a constructive derivation of the various expressions.

Note that for small values of $\sqrt{c} |x|$ ($ < 1.3 $) the
curvature becomes positive, while for larger values it is
negative, achieving anti-de Sitter space at spatial infinity.
Similar behavior is seen in the 3-dimensional curvature, which
according to (\ref{Rred}) is
\begin{equation}\label{Rredkink}
  R = - \frac{3 c}{2} + \frac{5 c}{2}
  \frac{1}{\cosh^2 \frac{\sqrt c}{2} x} \;.
\end{equation}
The corresponding 3-dimensional line element reads
\begin{equation}\label{ds3dimkink}
  \left( d S \right)^2 = - \left( d x \right)^2
  - 2 \frac{dt \; dy}{\cosh^2 \frac{\sqrt c}{2} x}  -
  \left( d y \right)^2 \;.
\end{equation}
Apart from $t$- and $y$- translations, the metric possesses no
other symmetries.

From (\ref{elf1}) we see that one may view $f$ as moving in a
potential
\begin{equation}\label{Vkink}
  V(f) = \frac{1}{4} (c - f^2)^2 \;.
\end{equation}
In flat space this potential also supports a kink with profile
\begin{equation}\label{flatkink}
  k (x) = \sqrt{c} \tanh \sqrt{\frac{c}{2}} x \;.
\end{equation}
Apart from a different scale on $x$, this coincides with our
general relativistic kink (\ref{fkink}). Note also that the
2-dimensional line element in (\ref{dskink}) may be presented in
terms of $V$, up to a rescaling of $t$
\begin{equation}\label{dskinkV}
\left( d s \right)^2 = V (f) \left( d t \right)^2 - \left( d x
\right)^2 \;.
\end{equation}

In Appendix B we show that these relations between general
relativistic and flat space kinks are not accidental, and can be
understood in general terms.

\section{Conclusions}

We have shown that dimensionally reducing the 3-dimensional
gravitational Chern-Simons term produces a 2-dimensional
topological theory. By itself it leads to symmetry breaking and
kink solutions, which structurally are very similar to certain
flat space kinks. Further investigation should address the
stability of time-dependent fluctuations around our static
solutions.

It is interesting that the kink causes the space to possess
asymptotic anti-de Sitter geometry, while carrying positive
curvature at small distances and vanishing at an intermediate
point. In this way the effect of the kink is analogous to a
geometric gravitational force, which has been previously explored
\cite{cangemi}: in 2-dimensional space-time one couples a
connection 1-form $A$ to the world line of a particle. $A$ is not
an external variable, but is implicitly constructed from
gravitational variables through the formula
\begin{equation}\label{cosmconst}
  \partial_\mu A_\nu - \partial_\nu A_\mu = \eps_{\mu \nu}
  \sqrt{- g} \Lambda \;,
\end{equation}
where $\Lambda$ is a constant. The covariantly constant 2-form,
$ d A$, causes the cosmological constant to shift by $2
\Lambda$ across the world line (which divides 2-dimensional
space-time). This mechanism also exists in higher dimensions. For
example in 4-dimensional space-time one couples a 3-form potential
to a membrane, and the ``field'' is a covariantly constant 4-form,
affecting the cosmological constant in a similar fashion
\cite{nicolai}.

Our 2-dimensional Lagrangian (\ref{cspiccola}) is formally similar to dilaton
Lagrangians. Indeed, when our $f$ is identified with the dilaton field,
(\ref{cspiccola}) becomes the $a = 0$, $b = 3$ special case in the class of dilaton models
discussed in \cite{kummer}, eq. (3.67). However, the crucial difference
is that $f$ is not a fundamental field, but a 2-dimensional curl of a
vector potential. In terms of the latter, the equations of motion are
of third derivative order, as is the Cotton tensor (\ref{cotton}). Evidently
our construction gives a suggestion for generalizing dilaton
theories \cite{kummer}: replace the dilaton scalar by the curl of a
vector.

Dimensional reduction of higher dimensional Chern-Simons terms
should also be interesting. A 5-dimensional Chern-Simons term
reduced to 4 dimensions should result in an action of the form
\begin{equation}\label{5dimcsred}
  \int d^4 x \; \Theta (R, F, \phi) \; {}^*F^{\mu \nu} F_{\mu \nu} \;,
\end{equation}
where the Chern-Pontryagin ``anomaly'' density ${}^* F^{\mu \nu}
F_{\mu \nu}$ is multiplied by the ``axion'' field $\Theta$, here
depending on the geometry $R$, the gauge field $F$, and a dilaton
$\phi$ [c.f. (\ref{cspiccolaxion})]. If $\Theta$ has strong
variation in only one direction of space-time, the effect of
(\ref{5dimcsred}) would be similar to the Chern-Simons
modification of Maxwell theory, which produces a Lorentz violating
Faraday rotation \cite{carroll1} (not seen experimentally
\cite{carroll2}). Reducing from higher dimensions could produce
non-Abelian gauge fields. Whether generalized kinks (solitons)
exist in these models remains an interesting but open question.

\vspace{.5cm}

\noindent {\bf \large Acknowledgments}

\noindent G.G. acknowledges hospitality at the Center for
Theoretical Physics - MIT, and A.I. thanks Jan Troost for
enjoyable discussions on 3-dimensional gravity. This work is
supported in part by funds provided by the US Department of Energy
(DOE) under cooperative research agreements DF-FC02-94ER40818,
DE-FG02-91ER40688-TaskD, and DE-FG02-91ER40676.

\newpage

\appendixa

We derive the kink profile (\ref{fkink}), together with
(\ref{rkink}) and (\ref{dskink}). The validity of these
expressions can be established by directly verifying that
(\ref{eljsol})-(\ref{elf2}) are satisfied\footnote{One would
compute $r$ and the covariant derivative from the line element
(\ref{dskink}).}. But it is interesting to give a construction.

We begin by using light-cone components $x^{\pm}$, and adopting
the 2-dimensional conformal gauge: $g_{\alpha \beta} = e^\sigma
\eta_{\alpha \beta}$, $\eta_{+ -} = 1$, $\eta_{+ +}= 0 = \eta_{-
-}$. First, we work on the consequences of (\ref{elf2}), which now
reads
\begin{equation}\label{a1}
\partial_\alpha (e^{-\sigma} \partial_\beta f ) +
\partial_\beta (e^{-\sigma} \partial_\alpha f ) =  \eta_{\alpha
\beta} \eta^{\gamma \delta} \partial_\gamma (e^{-\sigma}
\partial_\delta f ) \;.
\end{equation}
This is recognized as the (flat, 2-dimensional) conformal Killing
equation for the quantity $e^{-\sigma} \partial_\alpha f$. It is
solved by the statement that $e^{-\sigma} \partial_\pm f$ depends
only on $x^\mp$
\begin{equation}\label{a2}
  e^{-\sigma} \partial_\pm f = a_\mp (x^\mp) \;.
\end{equation}
It further follows that
\begin{equation}\label{a3}
 a_+ (x^+) \partial_+ f = e^\sigma a_+(x^+) a_-(x^-) =
 a_- (x^-) \partial_- f \;.
\end{equation}
When we introduce new variables $X^\pm$ through the relation
\begin{equation}\label{a4}
  X^\pm = \int^{x^\pm} \frac{d u}{a_\pm (u)} \;,
\end{equation}
Eq.(\ref{a3}) reads
\begin{equation}\label{a5}
  \frac{\partial}{\partial X^+} f = \frac{\partial}{\partial X^-} f
  = e^\sigma a_+  a_- \;,
\end{equation}
or, going to the Cartesian basis\footnote{In spite of the
notation, we are not assuming that $T$ is a time-like coordinate
and $X$ space-like. The identification will emerge presently.}
\begin{equation}\label{a6}
  X^\pm = \frac{1}{\sqrt 2}(X \pm T) \;,
\end{equation}
Eq.(\ref{a5}) states that $f$ is $T$-independent, and depends only
on $X$. From (\ref{a5}) it also follows that
\begin{equation}\label{a7}
  e^\sigma a_+ a_- = \frac{1}{\sqrt 2} \frac{d f}{d X} \;.
\end{equation}
Consequently, the line element may be written as
\begin{eqnarray}
\left( d s \right)^2 &= &e^\sigma 2 d x^+ d x^- = e^\sigma \frac{d
x^+}{d X^+} \frac{d x^-}{d X^-} 2  d X^- d X^- \nonumber \\
&=& \frac{1}{\sqrt 2}  \frac{d f}{d X} \left[ \left( d X \right)^2
- \left( d T \right)^2 \right] \label{a8} \;.
\end{eqnarray}

Next we turn to (\ref{eljsol}) and (\ref{elf1}). In our initial
conformal light-cone coordinates these read
\begin{eqnarray}
2 e^{-\sigma} \partial_+ \partial_- \sigma & = & c - 3 f^2
\label{a9} \;, \\
2 e^{-\sigma} \partial_+ \partial_- f & = & c f - f^3 \label{a10}
\;.
\end{eqnarray}
With (\ref{a4}), one converts them to
\begin{eqnarray}
2 \frac{e^{- \sigma}}{a_+ a_-} \; \frac{\partial}{\partial X^+}
 \frac{\partial}{\partial X^-} \ln  \frac{\partial f}{\partial X}
& = & c - 3 f^2
\label{a11} \;, \\
2 \frac{e^{- \sigma}}{a_+ a_-} \; \frac{\partial}{\partial X^+}
 \frac{\partial}{\partial X^-} f & = & c f - f^3 \label{a12} \;.
\end{eqnarray}
In (\ref{a9}), (\ref{a7}) was used to reexpress $
\frac{\partial}{\partial X^+} \frac{\partial}{\partial X^-}
\sigma$. We thus arrive at
\begin{eqnarray}
\frac{\sqrt 2}{d f / d X} \; \frac{d^2}{d X^2} \ln \frac{d f}{d X}
& = & c - 3 f^2
\label{a13} \;, \\
\frac{\sqrt 2}{d f / d X} \; \frac{d^2}{d X^2} f & = & c f - f^3
\label{a14} \;.
\end{eqnarray}
Observe that (\ref{a13}) is a consequence of (\ref{a14})
(differentiate the latter and divide by $d f / d X$ to obtain the
former). So we need not consider (\ref{a13}) any further and
concentrate on (\ref{a14}), which reads
\begin{equation}\label{a15}
  \sqrt{2} \; \frac{d^2}{d X^2} f =  (c f - f^3)  \frac{d f}{d X}
  \;.
\end{equation}
The first integral is immediate
\begin{equation}\label{a16}
  \sqrt{2} \; \frac{d f}{d X}  = - \frac{1}{4} (c  - f^2)^2 \;,
\end{equation}
where the integration constant is chosen so that, as $\left| X
\right| \to \infty$ and $f^2 \to c$, $d f / d X$ vanishes. The
next integral gives
\begin{equation}\label{a17}
  - \frac{1}{\sqrt 2} c^{3/2} X = \frac{2 \sqrt{c} f}{c - f^2}
  + \ln \left| \frac{\sqrt{c} + f}{\sqrt{c} - f} \right| \;.
\end{equation}

To proceed, we change variables from $X$ to $z$, through the
formula
\begin{equation}\label{a18}
  f (X) = \sqrt{c} \tanh m z \;,
\end{equation}
where $m$ ia an arbitrary parameter. From (\ref{a17}), this
implies that
\begin{equation}\label{a19}
  - \frac{1}{\sqrt 2} c^{3/2} X = 2 m z + \sinh 2 m z \;.
\end{equation}
We rewrite the line element (\ref{a8}) in terms of the new
coordinate
\begin{equation}
\left( d s \right)^2 = \frac{1}{\sqrt 2} \frac{\sqrt{c} m}{\cosh^2
m z} \frac{d z}{d X} \left[ \left( \frac{d X}{d z} \right)^2
\left( d z \right)^2- \left( d T \right)^2 \right] \label{a20} \;.
\end{equation}
The remaining derivative is evaluated from (\ref{a19})
\begin{equation}\label{a21}
  - \frac{1}{\sqrt 2} c^{3/2} \frac{dX}{dz} = 4 m \cosh^2 m z \;,
\end{equation}
and the line element (\ref{a20}) becomes
\begin{equation}
\left( d s \right)^2 = \frac{c^2}{8} \frac{1}{\cosh^4 m z}  \left(
d T \right)^2 - \frac{4 m^2}{c} \left( d z \right)^2  \label{a22}
\;.
\end{equation}
A final rescaling
\begin{equation}\label{a23}
  T = \frac{\sqrt 8}{c} \bar{t} \;, \quad z = \frac{\sqrt c}{2 m}
  \bar{x} \;,
\end{equation}
arrives at the line element (\ref{dskink}), and the kink profile
(\ref{a18}) agrees with (\ref{fkink}).

Note that the disposition of signs in the line element shows that
$T$ is a time-like variable, and $X$ space-like: our kink is
static, a derived result which was not assumed {\it a priori}.


\appendixb

We derive a general relation between flat space and general
relativistic kinks. A specific instance of this relations is
provided by the model discussed in the text.

With a specific potential $V$, see (\ref{Vkink})-(\ref{dskinkV}),
the flat space kink, $k$, depending on a single spatial variable,
satisfies the equation
\begin{equation}\label{b1}
 k'' = V' (k) \;,
\end{equation}
with first integral
\begin{equation}\label{b2}
  k' = \sqrt{2 V (k)} \;,
\end{equation}
where prime indicates differentiation with respect to the
argument.

We shall consider general relativistic kinks obeying
\begin{eqnarray}\label{b3}
  D^2 f & = & - V' (f) \;, \\
  D_\alpha D_\beta f & = & \frac{1}{2} g_{\alpha \beta} D^2 f
  \label{b4} \;.
\end{eqnarray}
As explained in Appendix A, Eq.(\ref{b4}) implies that $f$ depends
only on $X$; this allows rewriting the left side of (\ref{b3}) so
that $f$ satisfies the equation
\begin{equation}\label{b5}
  \frac{\sqrt 2}{d f / d X} \frac{d^2}{d X^2} f = - V' (f) \;,
\end{equation}
and leads to the line element
\begin{equation}\label{b6}
\left( d s \right)^2 = \frac{1}{\sqrt 2} \frac{d f}{d X} \left[
\left( d X \right)^2 - \left( d T \right)^2 \right]  \;,
\end{equation}
[compare (\ref{a8}) and (\ref{a14})].

Equivalent to (\ref{b5}) is
\begin{equation}\label{b7}
  \sqrt{2} \frac{d^2}{d X^2} f = - V'(f) \frac{d f}{d X} \;,
\end{equation}
with immediate first integral
\begin{equation}\label{b10}
\frac{d f}{d X} = - \frac{1}{\sqrt 2} V(f) \;.
\end{equation}
The second integral gives
\begin{equation}\label{b11}
  \sqrt{2} \int^{f(X)} \frac{d u}{V(u)} = - X \;.
\end{equation}

Next, we change variables from $X$ to $z$, using the flat space
kink $k$ of (\ref{b1}), and (\ref{b2}) arising from the same
potential $V$ [c.f. (\ref{a18})]
\begin{equation}\label{b12}
  f (X) = k (z) \;,
\end{equation}
\begin{equation}\label{b13}
  \sqrt{2} \int^{k (z)} \frac{d u}{V(u)} = - X \;.
\end{equation}
The line element (\ref{b6}) is now rewritten as
\begin{eqnarray}
  \left( d s \right)^2 & = & \frac{1}{2} V (f) \left( d T \right)^2
  - \frac{1}{\sqrt2} \frac{d k(z)}{d z} \frac{d z}{d X}
  \left[ \left( \frac{d X}{d z} \right)^2 \left( d z \right)^2
  \right]\nonumber \\
  &=& \frac{1}{2} V (f) \left( d T \right)^2
  - \frac{1}{\sqrt2} \frac{d k(z)}{d z} \frac{d X}{d z}
  \left( d z \right)^2 \label{b14} \;.
\end{eqnarray}
The derivative of $X$ is evaluated from (\ref{b13})
\begin{equation}\label{b15}
  \frac{d X}{d z} = - \frac{\sqrt2}{V(k)} \frac{d k(z)}{d z} \;,
\end{equation}
giving for the line element
\begin{equation}\label{b16a}
  \left( d s \right)^2 = \frac{1}{2} V (k) \left( d T \right)^2
  - \frac{1}{V(k)} \left( \frac{d k(z)}{d z} \right)^2
  \left( d z \right)^2 \;.
\end{equation}
But, from (\ref{b2}) this is
\begin{equation}\label{b16b}
  \left( d s \right)^2 = \frac{1}{2} V (k) \left( d T \right)^2
  - 2 \left( d z \right)^2 \;.
\end{equation}
A final rescaling $z = \frac{1}{\sqrt2} \bar{x}$, $T = \sqrt{2}
\bar{t}$, gives agreement with (\ref{dskinkV}), and shows from
(\ref{b12}) that general relativistic kink coincides with
flat-space kink, apart from rescaling of the argument
\begin{equation}\label{b18}
  f = k \left( X / \sqrt{2}\right) \;.
\end{equation}

The 2-curvature associated with the geometry (\ref{dskinkV}) is
\begin{equation}\label{b19a}
  r = \frac{1}{2} \frac{1}{V^2} \left( \frac{d V}{d X} \right)^2
  - \frac{1}{V} \frac{d^2 V}{d X^2} \;.
\end{equation}
The derivatives are evaluated by recalling that $V$ depends on the
rescaled kink profile (\ref{b18})
\begin{eqnarray}
\frac{d V}{d X} & = & \frac{1}{\sqrt2} V' k' = V' \sqrt{V}
\label{b20} \;, \\
\frac{d^2 V}{d X^2} & = & \frac{k'}{\sqrt2} \left( V'' \sqrt{V} +
\frac{1}{2} \frac{(V')^2}{\sqrt V} \right) = V'' V + \frac{1}{2}
(V')^2 \label{b21} \;.
\end{eqnarray}
In the last equality we used (\ref{b2}). Thus
\begin{equation}\label{b22}
  r = - V'' \;.
\end{equation}

All these relations are realized with the specific example considered
in the text, where $V = \frac{1}{4} (c - f^2)^2$, compare
(\ref{eljsol}), (\ref{fkink}), (\ref{dskink}), (\ref{flatkink}).

\end{document}